\newif\ifpretty
\prettytrue

\ifpretty
\documentclass[pra,twocolumn,amsmath,amssymb,a4paper,twoside,superscriptaddress]{revtex4}
\else
\documentclass[pra,preprint,amsmath,amssymb,a4paper,superscriptaddress]{revtex4}
\fi

\usepackage{graphicx}
\DeclareGraphicsExtensions{.eps}
\graphicspath{{./FIG/}}

\usepackage{dcolumn}

\begin{document}

 
\title{Adding Salt to an Aqueous Solution of t-Butanol: Is 
  Hydrophobic Association
  Enhanced or Reduced?}

\author{Dietmar Paschek} 
\email{dietmar.paschek@udo.edu}
\homepage{http://ganter.chemie.uni-dortmund.de/~pas}
\thanks{Fax: +49-231-755-3937}
%
\author{Alfons Geiger} 
\affiliation{Physikalische Chemie und}
\author{Momo Jeufack Herv\'e} 
\author{Dieter Suter} 
\affiliation{Fachbereich Physik, Universit\"at Dortmund,
           D-44227 Dortmund, Germany}

\date{\today}
\begin{abstract}
Recent neutron scattering experiments on aqueous salt 
solutions of amphiphilic t-butanol by Bowron and Finney 
[Phys. Rev. Lett. {\bf 89}, 215508 (2002); 
J. Chem. Phys. {\bf 118}, 8357 (2003)]
suggest the formation of t-butanol pairs, bridged by a
chloride ion via $\mbox{O}\!-\!\mbox{H}\cdots\mbox{Cl}^-$
hydrogen-bonds, and leading to a reduced
number of intermolecular hydrophobic butanol-butanol contacts. 
Here we present a joint experimental/theoretical study 
on the same system, using a 
combination of molecular dynamics simulations 
and nuclear magnetic relaxation measurements. 
Both MD simulation and experiment clearly 
support the more classical scenario of an
enhanced number of hydrophobic contacts in the presence
of the salt, as it would be expected 
for purely hydrophobic solutes [J. Phys. Chem. B {\bf 107}, 612 (2003)]. 
Although our conclusions arrive at a structurally completely distinct
scenario, the molecular dynamics simulation results are within the
experimental errorbars of the Bowron and Finney work.
\end{abstract}

\maketitle

\section{INTRODUCTION}

Nonpolar solutes, such as noble gases or alkanes, don't like to be
dissolved in water. Consequently, they are considered as ``hydrophobic''
and their corresponding
solvation free energy is found to be
large and positive \cite{Tanford,Ben-Naim:Hydrophobic,Pratt:2002:1,Southall:2002,Widom:2003}.
This effect is typically found to be significantly strengthened when salt is added,
leading to a further reduced solubility of hydrophobic species
such as noble gases, or methane \cite{Masterton:1971,Kinoshita:97}. The increasing excess chemical
potential is usually found to be proportional to the salt concentration 
over large concentration ranges and is therefore parameterized in
terms of Setschenow's concentration independent salting out coefficient \cite{Masterton:1971}.
In line with the observation of an increased positive solvation free
energy upon addition of salt, Ghosh et al. \cite{Ghosh:2003} report
an increased number of hydrophobic contacts in a diluted aqueous solution
of methane. Moreover, for the case of hydrophobic interactions
in a hydrophobic polymer chain, an approximately linear relationship
between the salt concentration and the strength of
pair-wise hydrophobic interactions has been determined \cite{Ghosh:2005}. This 
observation seems to be in line with the general finding of Koga et al., that the
excess chemical potential of a small hydrophobic particle and the strength of 
hydrophobic pair interactions appears to be (almost) linearly related
\cite{Widom:2003,Koga:2004}.

However, purely hydrophobic compounds are probably 
not very typical representatives of biophysical constituents. 
Usually, proteins and membrane-forming lipids
 are amphiphilic in the sense that they are composed
of both, hydrophobic and hydrophilic groups, where the latter ones
ensure a sufficiently high solubility in an aqueous environment.
In addition, the delicate interplay between hydrophobicity and hydrophilicity is
exploited by nature to control the dimensions
of molecular aggregates in aqueous solution
\cite{Huang:2000,Ashbaugh:2005,Rajamani:2005,Chandler:NatureReview}.
Small amphiphilic alcohols might thus be considered as 
a minimalist model system to explore the subtle interplay between hydrophobic
and hydrophilic effects. Numerous experimental studies on alcohol aggregation 
in aqueous solution have been reported, based on nuclear magnetic resonance
\cite{Holz:1992,Holz:1993,Sacco:1998,Mayele:1999,Mayele:2000},
as well as light-, x-ray-, and neutron scattering techniques
\cite{Euliss:1984,Bowron:1998:1,Bowron:2001,Bowron:2002,Dixit:2002,Bowron:2002:1,Bowron:2003}.

Salts are known to influence a number of properties
of aqueous solutions in a systematic way \cite{Cacace:1997}. The effect of
different anions and cations appears to be ordered in a sequence,
already proposed by Hofmeister in 1888 \cite{Hofmeister:1888}, 
deduced from a series of
experiments on the salts ability to precipitate ``hen-egg white protein''.
However, the exact reason for the observed specific cation and anion sequences
is still not completely understood, since
the same salt that can precipitate a protein at one concentration can
'salt it in' at another \cite{Parsegian:1995}.
Model calculations \cite{Hribar:2002}, as well as nuclear magnetic relaxation
experiments \cite{Holz:1993} propose a delicate balance between 
ion adsorption and exclusion  at the solute interface, tuned by the
solvent (water) structure modification according to 
the ion hydration \cite{Geiger:81,Leberman:1995} and hence possibly
subject to molecular details.

Recently, Bowron and Finney provided a detailed mechanistic picture of the
possible salting out process of t-butanol (TBA) in aqueous solution. 
The atomistic structure of the solution was 
determined from neutron scattering experiments
\cite{Bowron:2002:1,Bowron:2003}, varying solute and solvent isotopic
compositions \cite{Finney:1994}. 
However, their analysis relies largely on 
the accuracy of the employed empirical potential structure refinement 
(EPSR) technique of Soper \cite{Soper:1996,Soper:2001}.
Their main observation is that TBA-molecules form dimers that
are connected by hydrogen bonds to a central chloride anion.
Salting out appears hence due to solute aggregates, which are
formed by anion-bridges between the hydroxyl-groups,
increasing the solutes overall hydrophobic surface and thus reducing
the solubility of the whole complex 
\cite{Bowron:2004,Finney:2004:1,Finney:2004:2}. 
A straightforward conjecture would suggest that 
the salting out of proteins
could be driven by analogous anion-bridged aggregates.
We would like to point out that the proposed mechanism has 
remarkable similarity with the ``differential hydrophobicity''
concept of Burke et al. \cite{Burke:2003} used to explain the
specific protein-aggregation behavior observed in Huntington's disease.

The molecular dynamics simulations discussed here,
however, do not show any evidence for a ``salting out''-scenario
as proposed by Bowron and Finney. Instead, upon addition of
salt we find an increased number of hydrophobic contacts of the
TBA molecules which increases with higher salt concentration.
Using a combination of molecular
dynamics simulations and nuclear magnetic relaxation experiments
we show that an association parameter based on NMR measurable
quantities and introduced by H.G. Hertz et al.
\cite{Hertz1967, Hertz1976} 
is a useful measure for the association 
of TBA molecules in the present case. 
Both simulation and NMR experiments consistently support
the classical picture of an 
enhancing hydrophobic association in the case
of aqueous TBA/salt solutions.

\section{METHODS}

\subsection{Dipolar nuclear magnetic relaxation and correlations in the 
structure and dynamics of aqueous solutions}

The molecular dynamics simulations yield the time-dependent 
positions of the atomic nuclei. 
Experimentally, the individual molecular positions are not available.
However, a useful measure of molecular association is experimentally
accessible via the measurement of nuclear spin relaxation rates
and is discussed in a separate section below. In this section we
would like to briefly summarize the underlying theory.

The most important contribution to the relaxation rate of nuclear
spins with $I = 1/2$ is the magnetic dipole-dipole interaction.
The relaxation rate, i.e. the rate at which the nuclear spin
system approaches thermal equilibrium, is determined by the time dependence
of the magnetic dipole-dipole coupling.
For like spins, it is \cite{Abragam1961}
\begin{eqnarray}\label{eq:relax}
T_1^{-1} & = &
2\gamma^4\hbar^2I(I+1)(\mu_0/4\pi)^2
\\
& &
\left\{
\int\limits_0^\infty 
\left< \sum_j^N
\frac{D_{0,1}[\Omega_{ij}(0)]}{r_{ij}^3(0)}
*
\frac{D_{0,1}[\Omega_{ij}(t)]}{r_{ij}^3(t)}
\right> e^{i\omega t} dt
\right. \nonumber\\
&+&4\left.
\int\limits_0^\infty 
\left< \sum_j^N
\frac{D_{0,2}[\Omega_{ij}(0)]}{r_{ij}^3(0)}
*
\frac{D_{0,2}[\Omega_{ij}(t)]}{r_{ij}^3(t)}
\right> e^{i2\omega t} dt
\right\}, \nonumber
\end{eqnarray}
where $D_{k,m}[\Omega]$ is the $k,m$-Wigner 
rotation matrix element of rank $2$.
The Eulerian angles $\Omega(0)$ and $\Omega(t)$ at time zero and time $t$
specify the dipole-dipole vector relative to the laboratory fixed frame
of a pair of spins, $r_{ij}$ denotes their separation distance and
$\mu_0$ is the permittivity of free space.
The sum indicates summation of all $j$ interacting like spins
in the entire system.
$\left<\ldots\right>$ denotes averaging over all equivalent nuclei $i$
and all time zeros.
For the case of an isotropic
fluid and in the extreme narrowing limit Eq.~(\ref{eq:relax})
simplifies to \cite{Westlund:1987}
\begin{eqnarray}\label{eq:Westlund:1987}
T_{1}^{-1}
& = &
2 \gamma^4 \hbar^2
I(I+1)
\left(\frac{\mu_0}{4\pi}\right)^2
\int\limits_0^\infty
G_2(t) \,dt\;.
\end{eqnarray}
The dipole-dipole correlation function here is
abbreviated as $G_2(t)$ and is available through
\cite{Westlund:1987,Odelius:1993}
\begin{eqnarray}\label{eq:dipolcor}
G_2(t) & = & \left< \sum_{j} r_{ij}^{-3}(0)\,r_{ij}^{-3}(t)
P_2\left[\,\cos \theta_{ij}(t) \right] \right>\;,
\end{eqnarray}
where
$\theta_{ij}(t)$ is the angle between the vectors $\vec{r}_{ij}$
joining spins $i$
and $j$ at time $0$ and at time $t$ \cite{Westlund:1987} and
$P_2$ is the second Legendre polynomial. 

To calculate the integral, the correlation function $G_2(t)$
can be separated
into an $r^{-6}$-prefactor, which is sensitive to
the structure of the liquid (average internuclear distances)
and a correlation time $\tau_2$, which is obtained as the time-integral
of the normalized correlation function $\hat{G}_2(t)$, and which is sensitive to the mobility of
the molecules in the liquid,
\begin{eqnarray}\label{eq:ddcorel}
\int\limits_0^\infty G_2(t) \,dt & = & \left<
  \sum_j r^{-6}_{ij}(0)
\right> \tau_2\;\,.
\end{eqnarray}

From MD-simulation trajectory data
the correlation function $G_2$ and hence $T_1$
can be calculated directly.
From the definition of the dipole-dipole correlation function in 
Eq.~(\ref{eq:dipolcor}) it follows
that the relaxation time $T_1$
is affected  by both, reorientational and translational
motions in the liquid. Moreover, it is obvious that it also
depends strongly on the average 
distance between the spins
and is hence sensitive to changing inter- and intramolecular
pair distribution functions \cite{Hertz1967, Hertz1976}. 
In addition, the $r^{-6}$-weighting introduces a particular
sensitivity to changes occurring at short distances.
For convenience, one may divide the spins $j$ into different
classes according to whether they belong to the same molecule as
spin $i$, or not, thus arriving at an {\em inter}- and {\em
intramolecular} contribution to the relaxation rate
\begin{eqnarray}
T_1^{-1} = T_{1,\rm inter}^{-1} + T_{1,\rm intra}^{-1} ,
\end{eqnarray}
which are determined by corresponding intra- and intermolecular
dipole-dipole correlation functions
$G_{2,\rm intra}$ and $G_{2,\rm inter}$.
The  intramolecular contribution is basically due
to molecular reorientations and conformational changes
and has been used extensively to study the reorientational
motions, such as that of the H-H-vector in $\mbox{CH}_3$-groups in
molecular liquids and crystals \cite{Stejksal:1959}.
In the course of this paper, however, we are particularly
interested in the association of solute molecules, and will therefore focus
on the intermolecular contribution (see also experimental section).

The structure of the liquid can be expressed in terms
of the intermolecular site-site pair correlation function $g_{ij}(r)$, 
describing the probability of finding
an atom of type $j$ in a distance r from a reference site 
of type $i$ according to \cite{Egelstaff}
\begin{equation}
g_{ij}(r) = \frac{1}{N_i\,\rho_j} 
\left< 
\sum_{k=1}^{N_i}
\sum_{l=1}^{N_j}
\delta(\vec{r}-\vec{r}_{kl})
\right>\;,
\end{equation}
where $\rho_j$ is the number density of atoms of type $j$.
The prefactor  of the intermolecular dipole-dipole correlation
function is hence related to the pair distribution function via
an $r^{-6}$ integral of the pair correlation function
\begin{equation}\label{eq:r6gr}
\left<\sum_j r^{-6}_{ij}(0)\right> = \rho_j 
\int\limits_0^\infty r^{-6} \; g_{ij}(r)\;4\pi\,r^2\,dr\;.
\end{equation}
Since the process of enhanced association in a molecular solution 
is equivalent with an increase of the nearest neighbor peak
in the radial distribution function, Eq.~\ref{eq:r6gr} 
establishes a quantitative relationship between the degree of
intermolecular association and the intermolecular dipolar
nuclear magnetic relaxation rate.

\subsection{Self-association: the Hertz ``A''-parameter}
\label{sec:aparameter}

As a measure of the degree of intermolecular
association, Hertz and co-workers \cite{Hertz1967, Hertz1976, Capparelli1978}
introduced a so-called ``association parameter $A$'', 
which is a weighted integral of the pair correlation
function of the nuclei contributing to the dipolar relaxation process
(in the present case $^1\mbox{H}$ nuclei in TBA-d1-solutions)
and is defined as \cite{Muller:1996}
\begin{equation}\label{eq:A_para_1}
    A=\frac{1}{2} 
    \frac{\gamma^4 \hbar^2}{a^4 }  
    \left(\frac{\mu_0}{4\pi}\right)^2
    \int_0^\infty \left(\frac{a}{r}\right)^6 
    \,g_{\rm HH}(r) \;4\pi r^2\;dr \;,
\end{equation}
where $a$ is the ``closest approach distance of the interacting nuclei''
and is usually assumed to be independent of the system's composition \cite{Muller:1996}.

The correlation time $\tau_2$ of the intermolecular dipole-dipole interaction
is approximated by \cite{Muller:1996}
\begin{equation}
\tau_{2,\rm inter} = \frac{a^2}{3\,D}\;,
\label{eq:A_approx}
\end{equation}
where $D$ is the self diffusion coefficient of the solute molecules. $D$ can
also be measured by NMR, using e.g. pulsed field gradient experiments \cite{Price1998}.

Changes of the $A$-parameter indicate short-range changes in the pair
correlation function, which in the present study characterize the
solvent-mediated interaction between the solute molecules. Enhanced
association is identified by an increasing $A$-parameter: As the
first neighbor peak of $g_{\rm HH}(r)$ becomes sharper, the $A$-parameter
increases

Using the definitions of Eq.~\ref{eq:A_para_1}
and Eq.~\ref{eq:A_approx}, $A$ is given 
in terms of NMR measurable quantities 
\cite{Wandle1992,Muller:1996}:
\begin{equation}\label{eq:A_para_2}
    A = \frac{1}{T_{1,\rm inter}}\times\frac{D}{\rho_{\rm H}}\;,
\end{equation}
where $\rho_{\rm H}$ is the number density of the
$^1\mbox{H}$-nuclei in the system. Note that in our study 
$\rho_{\rm H}$ is kept nearly constant, varying only the additional
salt content, whereas in most NMR-studies 
\cite{Holz:1992,Holz:1993,Sacco:1998,Mayele:1999,Mayele:2000,Muller:1996,Wandle1992}
the concentration of the molecules, whose aggregation behavior is studied,
is varied \cite{Mayele:1999}.

Besides detecting $T_{\rm 1,inter}$ and $D$ by NMR, we can also calculate
these quantities, which are required for $A$, independently from our simulations
and determine
the $A$-parameter for our model system exactly
the same way as it is done from the  experiment. 
Since the TBA-TBA pair-correlation function and the
corresponding coordination number is also
available from MD, the simulations thus
provide a ``proof of concept'' for a system
behaving closely similar to the real system.

\subsection{MD simulation details}
\label{sec:MD}

\begin{table}[!t]
\caption{
  Nonbonded interaction parameters used in the present study.
  The $^1$ refers to the ion parameter set of Heinzinger \cite{Heinzinger:90}, 
  whereas the $^2$ refers to the parameters of Koneshan et al.\
  \cite{Koneshan:1998}. Lorentz-Berthelot mixing rules 
  according to $\sigma_{ij}\!=\!(\sigma_{ii}+\sigma_{jj})/2$ and
  $\epsilon_{ij}\!=\!\sqrt{\epsilon_{ii}\,\epsilon_{jj}}$
  were employed.
}
\label{tab:PARAM}
\centering
  \renewcommand{\arraystretch}{1.1}
  \renewcommand{\tabcolsep}{0.52cm}
  \small
  \begin{tabular}{lccc} \hline\hline \\[-6pt]
Site & $q / |e|$ & $\sigma/\mbox{nm}$ & $\epsilon k^{-1}/\mbox{K}$ \\ \hline\\[-6pt]
OW  & -0.8476 & 3.1656 & 78.2 \\
HW  & +0.4238 & --     & --   \\[6pt]
CT  & +0.265  & 3.50   & 33.2 \\
CT($\mbox{CH}3$)  & -0.180 & 3.50 & 33.2 \\
HC      & +0.060  & 2.50   & 15.1  \\
OH      & -0.683  & 3.12   & 85.6  \\
HO      & +0.418  & --     & --    \\[6pt]
Na$^1$  & +1.0    & 2.73   & 43.06 \\
Cl$^1$  & -1.0    & 4.86   & 20.21 \\
Na$^2$  & +1.0    & 2.583  & 50.32 \\
Cl$^2$  & -1.0    & 4.401  & 50.32 \\[6pt] \hline\hline
  \end{tabular}
\end{table}
\begin{table}[!t]
\caption{
  Parameters characterizing the performed MD-simulation runs. 
  All simulations were
  carried out at $T\!=\!298\,\mbox{K}$ and $P\!=\!1\,\mbox{bar}$. 
  The star $^*$
  indicates the simulation run employing the parameters of Koneshan et
  al. \cite{Koneshan:1998} for $\mbox{NaCl}$. 
  For comparison a pure water simulation run of 500 SPCE molecules 
  over $20\,\mbox{ns}$  was performed yielding an average density
  $998.4\,\mbox{kg}\,\mbox{m}^{-3}$.
}
\label{tab:MDSIM}
\centering
  \renewcommand{\arraystretch}{1.1}
  \renewcommand{\tabcolsep}{0.2cm}
  \small
  \begin{tabular}{lccccc} \hline\hline \\[-6pt]
$N(\mbox{H}_2\mbox{O})$                                   & 1000   & 1000    & 1000   &  1000    \\
$N(\mbox{TBA})$                                           & 20     & 20      & 20     &    20    \\
$N(\mbox{NaCl})$                                          & --     & 10$^*$  & 10     &    20    \\
Simul. length  $\tau/\,\mbox{ns}$                         & 50     & 100     & 100    &  100     \\
Density    $\left<\rho\right>/\,\mbox{kg}\,\mbox{m}^{-3}$ & 991.5  & 1008.6  & 1002.6 & 1013.7   \\
$\left<c(\mbox{TBA})\right>/\,\mbox{mol}\;\mbox{l}^{-1}$  & 1.0170 & 1.0044  & 0.9984 & 0.9810   \\
$\left<c(\mbox{NaCl})\right>/\,\mbox{mol}\;\mbox{l}^{-1}$ & --     & 0.5022  & 0.4992 & 0.9810    
\\[6pt] \hline\hline
  \end{tabular}
\end{table}
We employ molecular dynamics (MD) simulations in the NPT ensemble using
the Nos\'e-Hoover thermostat
\cite{Nose:84,Hoover:85}
and the Rahman-Parrinello barostat
\cite{Parrinello:81,Nose:83} with
coupling times $\tau_T\!=\!1.5\,\mbox{ps}$ and
$\tau_p\!=\!2.5\,\mbox{ps}$
(assuming the isothermal compressibility to be
$\chi_T\!=\!4.5\;10^{-5}\,\mbox{bar}^{-1}$), respectively.
The electrostatic interactions are treated
in the ``full potential'' approach
by the smooth particle mesh Ewald summation
\cite{Essmann:95} with a real space
cutoff of $0.9\,\mbox{nm}$ and a mesh spacing of approximately
$0.12\,\mbox{nm}$ and 4th order
interpolation. The Ewald convergence factor $\alpha$ was set to
$3.38\,\mbox{nm}^{-1}$ (corresponding to a relative accuracy of
the Ewald sum of $10^{-5}$).
A $2.0\,\mbox{fs}$
timestep was used for all simulations and the solvent constraints were solved
using the SETTLE procedure \cite{Miyamoto:92}, while the SHAKE method
was used to constrain the solute bond lengths \cite{Ryckaert:77}.
All simulations reported here were carried out using
the GROMACS 3.2  program \cite{gmxpaper,gmx32}.
The MOSCITO suit of programs \cite{MOSCITO} was employed to 
generate start configurations, topology
files, and was used for the entire data analysis
presented in this paper.
Statistical errors in the analysis 
were computed using the method of Flyvbjerg and Petersen \cite{Flyvbjerg:89}.
For all reported systems 
initial equilibration runs of $1\,\mbox{ns}$ length
were performed using the Berendsen
weak coupling scheme for pressure and temperature control
($\tau_T\!=\!\tau_p\!=\!0.5\,\mbox{ps}$) \cite{Berendsen:84}.

As in Refs.~\cite{Bowron:2002:1,Bowron:2003} we study 0.02 mole
fraction aqueous solutions of t-butanol with and without 
presence of sodium chloride. The simulations
were carried out for $1\,\mbox{bar}$ and $298\,\mbox{K}$.
Our model system contains 1000 water molecules, represented
by the three center SPCE model \cite{Berendsen:87}. The flexible OPLS all-atom
forcefield \cite{Jorgensen:96} is employed for the 20 TBA-molecules.
Here the bond lengths were kept fixed.
Ten and twenty sodium chloride ion pairs were 
used to represent the salt solution. 
In order to check the influence of ion parameters variation,
the two different parameter sets according to Heinzinger \cite{Heinzinger:90}
and Koneshan et al.\ \cite{Koneshan:1998} were employed. All non-bonded
interaction parameters are summarized in Table~\ref{tab:PARAM}.
To ensure proper sampling, and to allow an accurate determination
of the system's structural and dynamical properties,
the aqueous TBA solutions were studied for $50\,\mbox{ns}$, whereas
the salt-solutions were monitored for $100\,\mbox{ns}$. The performed
simulation runs and resulting concentrations 
are indicated in Table~\ref{tab:MDSIM}. For comparison,  a pure water system
of 500 SPCE water molecules was simulated for $20\,\mbox{ns}$ for the
same conditions.

\subsection{Experimental Details}

\begin{table}[!b]
\caption{
  Experimental densities $\rho$, intermolecular relaxation times 
  $T_{1,\rm inter}$ and
  self-diffusion coefficients $D$ for TBA-d1 in 
  TBA-d10/$\mbox{D}_2\mbox{O}$/NaCl solutions.
  All experiments were
  carried out at $T\!=\!298\,\mbox{K}$ at ambient pressure 
  conditions.
  Also given are the TBA-d1 concentrations and the obtained
  $A$-parameters.
}
\label{tab:EXPT}
\centering
  \renewcommand{\arraystretch}{1.1}
  \renewcommand{\tabcolsep}{0.10cm}
  \small
  \begin{tabular}{lcccc} \hline\hline \\[-6pt]
TBA-d1:$\mbox{D}_2\mbox{O}$:NaCl    
& 2:100:0    & 2:100:1   & 2:100:2   \\
$\rho/\mbox{kg}\,\mbox{m}^{-3}$
& 1083.9    & 1106.1   & 1122.5   \\
$c(\mbox{TBA})/\mbox{mol}\,\mbox{l}^{-1}$
& 1.0059    & 0.9994   & 0.9882 \\
$c(\mbox{NaCl})/\mbox{mol}\,\mbox{l}^{-1}$
& -   & 0.5270  & 1.0403   \\
$T_{1,\rm inter} / \mbox{s}$       
& $44.89\pm1.74$ & $41.02\pm4.73$ &
$37.92\pm0.13$ \\
$D /10^{-9}\mbox{m}^2\mbox{s}^{-1}$   
& 0.3962 & 0.3818 & 0.3792 \\
$A /10^{-39}\mbox{m}^{5}\mbox{s}^{-2}$   
& 1.619 & 1.718 & 1.826
 \\[6pt] \hline\hline
  \end{tabular}
\end{table}
To experimentally determine the association behavior of TBA molecules
in aqueous solutions we measured 
intermolecular NMR relaxation rates,
self diffusion coefficients, and the densities of the aqueous solutions.
All experimental data are summarized in Table \ref{tab:EXPT}.

The observed
relaxation rates depend on inter- as well as intra-molecular
correlation functions. To extract the intermolecular rates, which 
are sensitive to the solute-solute
association, we used the method of
isotopic dilution \cite{Bonera1965}.

Since we are interested only in the hydrophobic methyl-protons, we
deuterated the water and the hydroxyl group of the TBA. Isotopic
dilution was performed by mixing (CD$_3$)$_3$COD (TBA-d10) with
(CH$_3$)$_3$COD (TBA-d1). We parameterize the dilution with the
mole fraction
$$
x_{\rm H} = \frac{[\mbox{TBA-d1}]}{[\mbox{TBA-d1}]+[\mbox{TBA-d10}]} .
$$
The basic assumption of the isotopic dilution procedure is that
the relaxation rate is given by the sum of an intramolecular term,
which is independent of the dilution, and an intermolecular term,
which is proportional to the concentration of the corresponding
molecular species. 
The contribution of the deuterated molecules can be taken as
proportional to that of the protonated molecules, with a reduction factor
\cite{Abragam1961}
$$
\alpha =    \frac{2}{3}  \frac{\gamma_D^2}{\gamma_H^2} \frac{I_D
(I_D+1)}{I_H (I_H+1)} =  0.042.
$$
The observed relaxation rate becomes therefore
\begin{equation}\label{eq.observed_rate}
        \frac{1}{T_{1}} = \frac{1}{T_{1,0}}+ \frac{1}{T_{1,\rm intra}}+
        \frac{1 }{T_{1,\rm inter}}[(1-\alpha) x_H + \alpha ].
\end{equation}
Here, $T_{1,\rm intra}$ denotes the intramolecular contribution from
protons within the methyl groups of the same molecule as the one
being measured, $T_{1,\rm inter}$ the intermolecular contributions
between different TBA molecules, and $T_{1,0}$ all other terms,
such as paramagnetic relaxation and interaction with other
molecules such as D$_2$O. To extract the intermolecular term, we
measured the relaxation rate as a function of the isotopic
dilution and fitted the measured data points to
Eq.~\ref{eq.observed_rate}.

The diffusion coefficients of TBA-d1
were determined from the Pulsed Gradient
Spin Echo (PGSE) experiments \cite{Price1998}, where the gradient
calibration was done using the diffusion coefficient of
pure water \cite{Mills1973}.

The $^1\mbox{H}$ number density in the 
TBA-d1/$\mbox{D}_2\mbox{O}$ and
TBA-d1/$\mbox{D}_2\mbox{O}$/NaCl solutions was obtained
by measuring the mass density of the 
corresponding solutions with defined
composition using a commercial
Anton Paar oscillating U-tube density meter.

The TBA-d1 ($99\%$) was purchased from Cambridge Isotope
laboratories, the TBA-d10 ($99\%$) from Isotec. The solvent D$_2$O
with the purity $99.96\%$ was obtained from Merck KGaA. The
solution was prepared by measuring the appropriate amount of each
compound with a micropipette, and by weighing a corresponding amount
of NaCl. The degassing process was done by the usual
Freeze-Pump-Thaw technique,  repeated several times until no gas
bubbles didn't develop from the solution. 
Finally, the samples were flame sealed. 
Relaxation and diffusion measurements were carried out at
$600\,\mbox{MHz}$ using a Varian Infinity Spectrometer system.
All experiments were conducted under controlled 
temperature conditions at $25^\circ\mbox{C}$.

\section{RESULTS AND DISCUSSION}

\subsection{Structural characterization of the aqueous TBA solutions}

\begin{figure}[!t]
  \centering
  \includegraphics[angle=0,width=4.5cm]{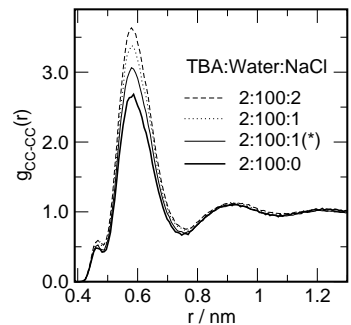}
  \caption{\footnotesize Radial pair distribution functions between the TBA central
    carbon atoms (CC) in aqueous solutions at different salt concentrations.
    The star $^*$ indicates the data 
    belonging to parameter set of Koneshan et al
    \cite{Koneshan:1998} for NaCl.
  }
  \label{fig:01}
\end{figure}

The most prominent feature of the 
combined neutron scattering/EPSR
work of Bowron and Finney is the observation of a
significant decrease of the height of the first peak of the
central-carbon
pair correlation function upon addition of 
sodium chloride \cite{Bowron:2002:1,Bowron:2003}. This 
decrease of the nearest neighbor peak is accompanied by an increase 
of the second neighbor peak located at a distance
of about $0.85\,\mbox{nm}$. From steric arguments and an analysis of
their EPSR data, Bowron and Finney conclude that this process is
according to the formation of chloride-bridged TBA-pairs.
In Figure \ref{fig:01} we present the corresponding curves 
obtained from the present MD simulations.
Surprisingly, exactly the opposite behavior 
is found. Upon addition of salt a notable increase of the
first peak is observed, suggesting an enhanced TBA-TBA aggregation. 
Moreover, this increase is clearly more pronounced
 when increasing the salt concentration
from about 0.5 molar to about 1 molar. Table~\ref{tab:COOR} 
provides a  quantitative analysis of the pair correlation 
data in terms of coordination numbers
\begin{eqnarray}
N_{\alpha\beta} & = & 4\pi \rho_\beta \int\limits_{r_{\rm min}}^{r_{\rm max}}
r^2g_{\alpha\beta}(r) \,dr\;,
\end{eqnarray}
where  $\rho_\beta$ is the average number density of atom type $\beta$.
In order to provide comparability with the data obtained by Bowron and
Finney, the values of $r_{\rm min}$ and $r_{\rm max}$ for the 
distance intervals given in the upper part of Table~\ref{tab:COOR} were 
taken from Ref.~\cite{Bowron:2003}. As the increasing peak height
of the MD-data suggests,
the TBA-TBA coordination number increases with increasing salt concentration.
Moreover, the increase is found to be significantly 
larger than the size of the associated errorbars.

\begin{figure}[!t]
  \centering
  \includegraphics[angle=0,width=5.0cm]{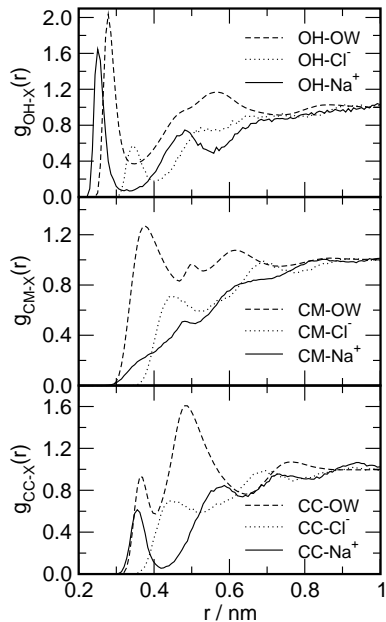}
  \caption{\footnotesize Several representative
    TBA-solvent pair atom-atom radial distribution  functions 
    obtained from the TBA:Water:NaCl 2:100:1 solution.
    CC denotes the TBA central carbon atom, whereas CM and OH specify
    the methyl-carbon and  hydroxyl oxygens, respectively.
  }
  \label{fig:02}
\end{figure}
\begin{figure}[!t]
  \centering
  \includegraphics[angle=0,width=4.0cm]{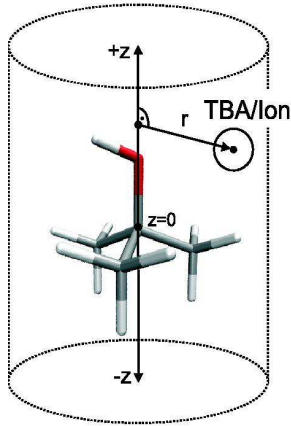}

  \caption{Schematic illustration of the two-dimensional cylindrical
    pair distribution functions. Distribution
    of TBA and the ions around a central TBA molecule. The center of
    mass of the central TBA molecules is at ($z\!=\!0$, $\!r=\!0$)
    and the C-O-bond is aligned along the $z$-axis.}
  \label{fig:03}
\end{figure}
\begin{table*}[!t]
\caption{Coordination numbers for a 0.02 molar aqueous solution
  of t-butanol (TBA) with and without added sodium chloride. The
  coordination numbers are obtained by integrating over the distance interval
  indicated. For a direct comparison, the values of
   Ref.~\cite{Bowron:2003} were taken
  for $r_{\rm min}$ and $r_{\rm max}$ for the upper part of the table.
  For the data shown in the lower part of the table (first shell coordination
  numbers) integration over $g(r)$
  was performed to the first minimum of the ion-oxygen pair correlation function.
  CC-refers to the central carbon atom of TBA. OW denotes the water-yxygen
  while OH specifies the oxygen atom in the hydroxyl-group. The star$^*$
  indicates the parameter set for NaCl according to Koneshan et
  al. \cite{Koneshan:1998}. The neutron scattering data were taken from
  Ref. \cite{Bowron:2003}. 
}
\label{tab:COOR}
\centering
  \ifpretty
  \renewcommand{\arraystretch}{1.1}
  \renewcommand{\tabcolsep}{0.12cm}
  \small
  \else
  \renewcommand{\tabcolsep}{0.02cm}
  \footnotesize
  \fi 
  \begin{tabular}{lcc|cc|ccc|c} \hline\hline \\[-6pt]
   & & & \multicolumn{6}{c}{TBA:Water:NaCl} \\
   & & & \multicolumn{2}{c}{2:100:0} &
\multicolumn{3}{c}{2:100:1} &
2:100:2
\\
  Atom-pair &
  $r_{\rm min}/\mbox{nm}$ &
  $r_{\rm max}/\mbox{nm}$ &
  n-scatt.&
  MD sim. &
  n-scatt.&
  MD sim. &
  MD sim.$^*$ &
  MD sim. 
\\[6pt] \hline \\[-6pt]
CC-CC&0.43 &0.75&$1.5\pm0.7$  &$1.15\pm0.01$  &$0.8\pm0.5$&$1.32\pm0.05$  &$1.27\pm0.01$  &$1.45\pm0.03$   \\
CC-CC&0.75 &1.00&$1.0\pm0.6$  &$1.41\pm0.02$  &$1.8\pm0.8$&$1.39\pm0.06$  &$1.39\pm0.02$  &$1.40\pm0.04$   \\
OH-OH&0.25 &0.35&$0.02\pm0.08$&$0.042\pm0.001$&\ldots     &$0.047\pm0.002$&$0.043\pm0.002$&$0.048\pm0.001$ \\
OH-OW&0.25 &0.35&$2.5\pm0.6$  &$3.017\pm0.002$&$2.4\pm0.6$&$2.975\pm0.005$&$2.992\pm0.002$&$2.950\pm0.002$   \\
Na-OW&0.21 &0.30&\ldots       &\ldots         &$4.2\pm1.1$&$5.562\pm0.004$  &$5.582\pm0.003$  &$5.454\pm0.002$   \\
Cl-OW&0.28 &0.36&\ldots       &\ldots         &$4.9\pm1.3$&$5.715\pm0.003$  &$6.319\pm0.003$  &$5.629\pm0.002$   \\
OH-Na&0.32 &0.55&\ldots       &\ldots         &$0.2\pm0.2$&$0.098\pm0.001$  &$0.101\pm0.001$  &$0.191\pm0.002$   \\
OH-Cl&0.29 &0.38&\ldots       &\ldots         &$0.2\pm0.2$&$0.0120\pm0.0002$ &$0.0134\pm0.0003$  &$0.0251\pm0.0004$  \\
HO-Cl&0.19 &0.32&\ldots       &\ldots         &$0.04\pm0.07$&$0.0135\pm0.0003$
   &$0.0141\pm0.0003$ &$0.0279\pm0.0004$ \\[2pt] \hline \\[-8pt]
Na-OH&0.0 &0.32&\ldots       &\ldots         & &$0.0307\pm0.0008$ &  &$0.0304\pm0.0008$   \\
Cl-OH&0.0 &0.39&\ldots       &\ldots         & &$0.0262\pm0.0005$ &  &$0.0275\pm0.0004$  \\
Na-OW&0.0 &0.32&\ldots       &\ldots         & &$5.780\pm0.004$   &  &$5.669\pm0.002$   \\
Cl-OW&0.0 &0.39&\ldots       &\ldots         & &$6.989\pm0.003$   &  &$6.940\pm0.002$  \\
OH-Na&0.0 &0.32&\ldots       &\ldots         & &$0.0154\pm0.0008$ &  &$0.0304\pm0.0008$   \\
OH-Cl&0.0 &0.39&\ldots       &\ldots         & &$0.0131\pm0.0005$ &  &$0.0275\pm0.0004$  \\
OW-Na&0.0 &0.32&\ldots       &\ldots         & &$0.0578\pm0.00005$   &  &$0.1134\pm0.00003$   \\
OW-Cl&0.0 &0.39&\ldots       &\ldots         & &$0.0699\pm0.00003$   &  &$0.1388\pm0.00004$  \\[6pt] \hline\hline
  \end{tabular}
\end{table*}
\begin{figure*}[!t]
  \centering
  \includegraphics[angle=0,width=13.5cm]{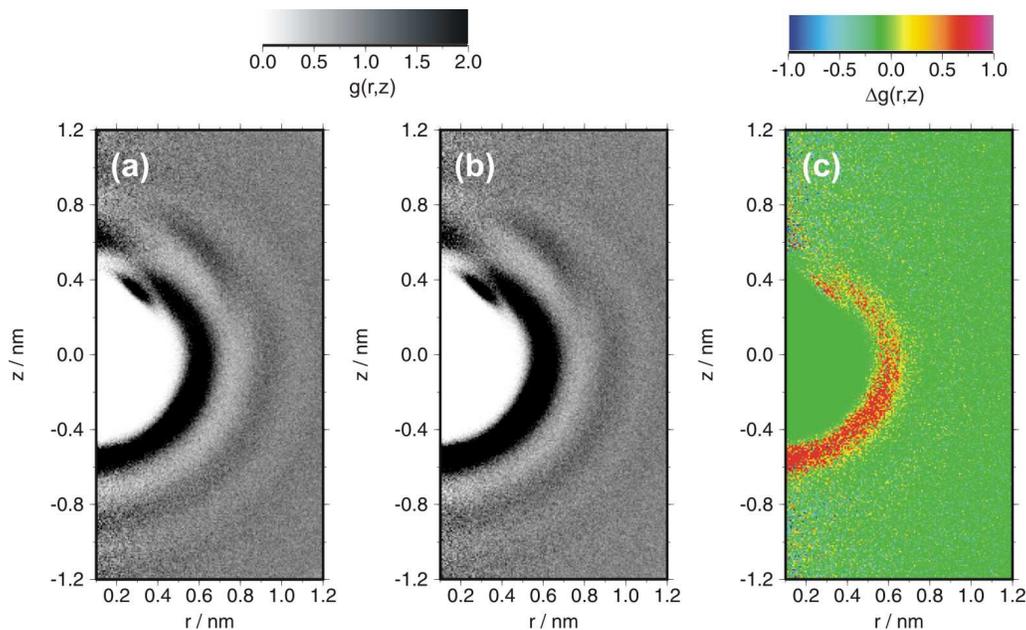}

  \caption{Two dimensional cylindrical center of mass pair distribution
    functions $g(r,z)$ of TBA around a central
    TBA-molecule with the C-O-Vector pointing upwards.
    (a) 2 TBA : 100 Water. (b) 2 TBA : 100 Water : 1 NaCl;
    (c) Difference between the distribution functions shown in (b) and (a)
    (salt minus no salt).
  }
  \label{fig:04}
\end{figure*}
\begin{figure}
  \centering
  \includegraphics[angle=0,width=8.5cm]{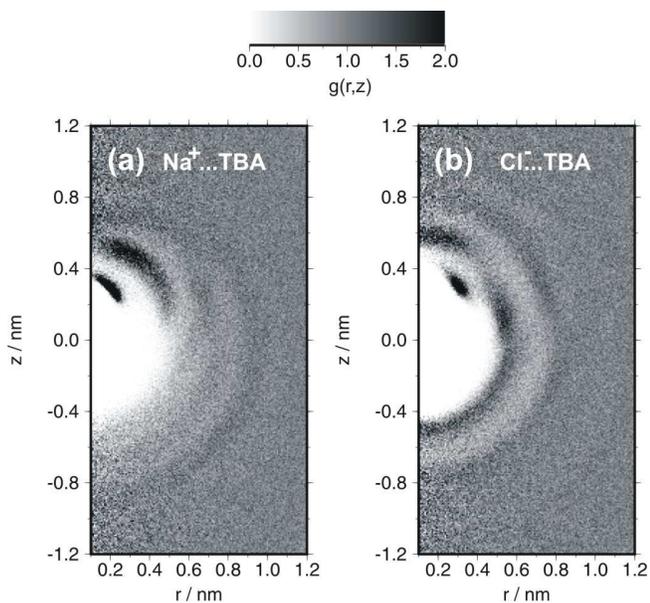}

  \caption{Two dimensional cylindrical pair distribution functions 
    $g(r,z)$ of the sodium (a) and chloride (b) ions around a  central
    TBA-molecule with the C-O-Vector pointing upwards. From
    the simulatio with composition TBA:Water:NaCl of 2:100:1.
  }
  \label{fig:05}
\end{figure}
\begin{figure}
  \centering
  \includegraphics[angle=0,width=4.0cm]{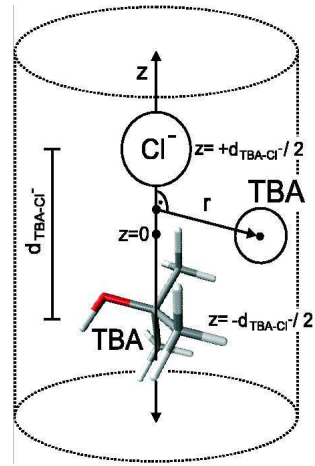}
  \caption{Schematic illustration of the two-dimensional cylindrical
    pair distribution function $g(r,z)$ of TBA molecules around a 
    $\mbox{TBA}-\mbox{Cl}^-$ contact pair with 
    $d_{\rm TBA-Cl^-}\!<\!0.64\,\mbox{nm}$.
    The origin ($z\!=\!0$,
    $\!r=\!0$) is located halfway between the
    centers of mass of $\mbox{TBA}$ and $\mbox{Cl}^-$.
    The vector connecting $\mbox{TBA}$ and $\mbox{Cl}^-$ is 
    aligned along the $z$-axis.
  }
  \label{fig:06}
\end{figure}
\begin{figure}
  \centering
  \includegraphics[angle=0,width=5.0cm]{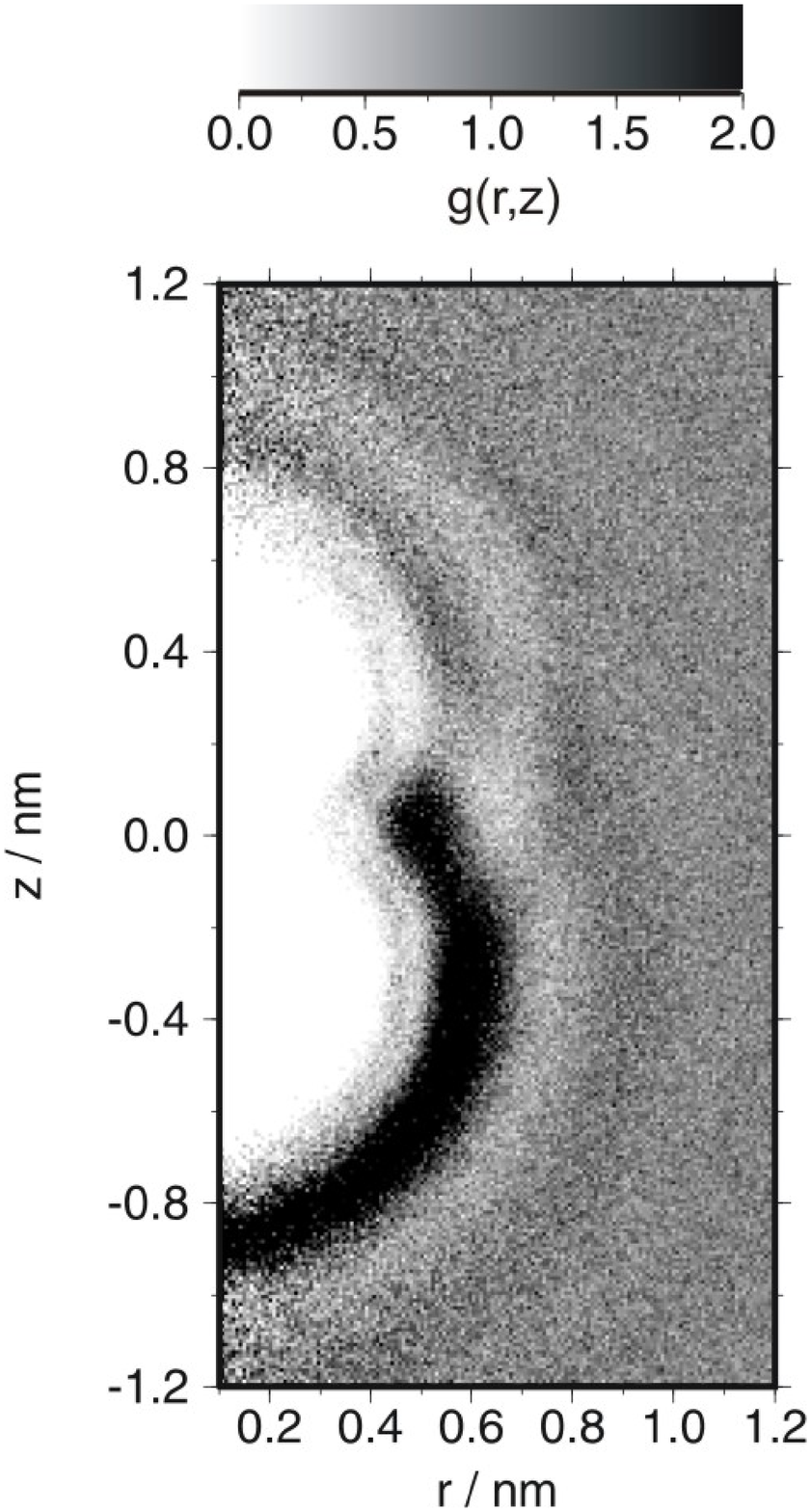}
  \caption{Cylindrical pair distribution function
    TBA around a
    contact pair formed by a TBA-molecule and a Cloride ion
    (TBA:bottom; $\mbox{Cl}^-$: top). The geometry and
    the definition of $r$ and $z$ is indicated
    in Figure \ref{fig:06}.
    }
  \label{fig:07}
\end{figure}

A more detailed picture of the structure of the aqueous salt solutions of
TBA is given in terms of selected solute/solvent pair correlation functions 
in Figure \ref{fig:02}. To quantify the change of ion-solvation with 
increasing salt concentration, ion/water-oxygen and ion/TBA-oxygen (first shell)
coordination numbers are given the in lower part of Table~\ref{tab:COOR}.
The changes in coordination numbers as the salt concentration doubles is only
relatively small. However, apparently there is a certain tendency of the ions 
to slightly dehydrate, whereas the Na-OH coordination number is almost unchanged
and the Cl-OH coordination slightly increases. 

In order to take the amphiphilic nature of TBA more properly into
account we also provide two-dimensional cylindrical pair correlation functions 
$g(z,r)$, indicating the arrangement of molecules around a central TBA
molecule using the notation of \cite{Egelstaff}
\begin{eqnarray}
g(r,z) & = & \frac{1}{N_{\rm TBA} \rho_\beta} 
\left<
\sum_{i=1}^{N_{\rm TBA}}
\sum_{j=1}^{N_{\beta}}
\delta\left(
z - \vec{n}_{\rm OH}\vec{r}_{ij}
\right)\right. \\\nonumber
& &
\left.
\times\; \delta\left(
r - \sqrt{%
r_{ij}^2- (\vec{n}_{\rm OH}\vec{r}_{ij})^2
}
\right)
\right>
\end{eqnarray}
where the $z$-axis is aligned along the TBA intramolecular 
unit-vector $\vec{n}_{\rm OH}\!=\!\vec{r}_{\rm OH}/r_{\rm OH}$ pointing from
the center of mass to the hydroxyl oxygen.
The index $i$ runs
over all TBA molecules, whereas the index $j$ runs over a particular subset of
molecules (TBA, anions, or cations). The 
vector $\vec{r}_{ij}\!=\!\vec{r}_{j}-\vec{r}_{i}$
represents the center of mass separation between particles $i$ and $j$.
The schematic shown in Figure \ref{fig:03} illustrates
how the parameters $r$ and $z$ are defined. 
The 2D-distribution of TBA molecules 
around a central TBA molecule is shown in Figure \ref{fig:04}. The radial pair
distribution functions in Figure \ref{fig:01} and the 2D-distributions
in Figure \ref{fig:04} are interrelated and the radial distribution functions
can be obtained by averaging over angles $\theta\!=\!\arctan\left(r/z\right)$.
Figure \ref{fig:04} reveals that the pre-peak in the CC-CC radial pair
distribution function
located a about $0.47\,\mbox{nm}$ is due to hydrogen bonded TBA-TBA
pairs. These pairs are identified by a separate 
dark spot at about $r\!=\!0.3\,\mbox{nm}$ and
$z\!=\!0.35\,\mbox{nm}$
in close proximity to the hydroxyl-group 
in Figures \ref{fig:04}a and \ref{fig:04}b.
As deduced from the radial distribution functions in Figure \ref{fig:01}, 
the addition of salt leads
to an increased aggregation of TBA. The difference between the two-dimensional
distribution functions with and without salt, shown in Figure \ref{fig:04}c, reveals that
aggregation occurs predominantly on the methyl-group side of the TBA
molecule: The region with negative $z$ shows an increase in peak height 
of about $0.5$, whereas on the hydrophilic 
side, the peak heights remain almost unchanged.

The distributions of the ions around a TBA-molecule are shown in
Figure \ref{fig:05}. Dark regions close to the hydroxyl group indicate
a significant stability of TBA-ion complexes. A remarkable 
difference is observed for adsorption of the different ions at the aliphatic
side of TBA.
The chloride ions tend to adsorb close to the Methyl-groups and
are present as a grey shadow on the hydrophobic side of the TBA 
molecule as shown in Figure \ref{fig:05}b, whereas the sodium 
completely tends to avoid that region. Consequently, there is practically no peak in
the methyl-carbon-sodium radial pair 
correlation function in Figure \ref{fig:02}. 
This observation seems to be in line with the recently
reported preferential anion-adsorption at the liquid-gas interface
\cite{Knipping:2000,Jungwirth:2000,Jungwirth:2001},
which has structural similarities with the interface to a hydrophobic surface
\cite{Huang:2000,Patel:2003,Brovchenko:2004}. The tendency of the chloride-ion
to attach to the methyl groups might also explain the observed slight increase in
the  Cl-OH coordination number. A chloride-ion just gets more frequently
attached to the OH-group since due to the preferential
methyl-group-chloride interaction the chloride sees a higher
local TBA concentration.

Finally, in order to prove whether there is a significant amount of 
chloride-bridged TBA-TBA configuration we calculate the two-dimensional 
cylindrical 
$(\mbox{TBA}-\mbox{Cl}^-)\ldots\mbox{TBA}$ correlation functions 
of a second TBA molecule around a $\mbox{TBA}-\mbox{Cl}^-$ contact pair
with a center of mass distance less than $0.64\,\mbox{nm}$
according to
\begin{eqnarray}
g(r,z) & = & \frac{1}{N_{\rm TBA-Cl}\;\rho_{\rm TBA}} \\ \nonumber
& &
\left<
\sum_{i=1}^{N_{\rm TBA-Cl}}
\sum_{j=1}^{N_{\rm TBA}} 
\delta\left(
z - \vec{n}_{\rm TBA-Cl}\vec{r}_{ij}
\right) \right. \\ \nonumber
& &
\left.
\times\; \delta\left(
r - \sqrt{%
r_{ij}^2- (\vec{n}_{\rm TBA-Cl}\vec{r}_{ij})^2
}
\right)
\right> \;.
\end{eqnarray}
Here $\vec{n}_{\rm TBA-Cl}$ is the unit vector describing the orientation
of a $\mbox{TBA}-\mbox{Cl}^-$ contact pair. An illustration is given
in Figure \ref{fig:06}. As can be seen clearly from Figure \ref{fig:07}, the
vast amount of TBA molecules is located on the TBA-side of the 
$\mbox{TBA}-\mbox{Cl}^-$-pair, practically ruling out  a significant
contribution of $\mbox{TBA}-\mbox{Cl}^--\mbox{TBA}$ bridges.

As the 2-dimensional cylindrical pair distribution functions indicate, 
the MD simulations furnish a scenario of an increased number of
hydrophobic contacts in an aqueous solution of TBA in the presence of sodium 
chloride, in contrast to the observations of Bowron and Finney.
However, when taking a look at the published data, and considering
the given size of errorbars, the n-scattering and MD simulation
are not at all contradictory.
A quantitative comparison of the
obtained coordination numbers is given  in Table \ref{tab:COOR}.
Almost all indicated values agree 
within the given errors, or are at least very close to each other.
In fact, the large error in the n-scattering data for
 the central carbon (CC) coordination numbers
for the first and second hydration shell, probably 
forbids a clear distinction
between both scenarios just relying on the n-scattering data.
\begin{table*}[!t]
\caption{
  Statistical analysis of the composition of the first solvation shell of
  the solvated Chloride-ion. $n$ specifies the number of TBA-molecules in
  the solvation shell. $P_n$ indicates the probability of finding a chloride
  ion with a solvation shell containing $n$ TBA molecules. 
  $\left<N_{\rm W}\right>_n$ indicates the average number of water molecules
  found in a solvation shell containing $n$ TBA molecules. The left part
  of the table indicates data obtained directly from the MD-simulations.
  In the right part of the Table the
  $P^{7(\rm id)}_n(K_{\rm eq})$ indicates the combinatorial prediction of the
  probability of finding a chloride ion with a solvation shell
  containing $n$ TBA molecules
  (see text for details). Here $K_{\rm eq}$ is the equilibrium constant
  describing the a priori Water/TBA equilibrium for each hydration shell site.
}
\label{tab:combinatoric}
  \centering
  \ifpretty
  \renewcommand{\tabcolsep}{0.58cm}
  \renewcommand{\arraystretch}{1.0}
  \else
  \renewcommand{\tabcolsep}{0.5cm}
  \renewcommand{\arraystretch}{1.0}
  \fi
 \begin{tabular}{l|cc|cc||ccc} \hline\hline \\[-4pt]
     &
   \multicolumn{4}{c}{MD Simulation:} &
   \multicolumn{3}{c}{Combinatorial Prediction:} \\[0.5em]
     &
   \multicolumn{2}{c}{2:100:1} &
   \multicolumn{2}{c}{2:100:2} \\[0.5em]
   $n$ & $P_n$ & $\left<N_{\rm W}\right>_n$ 
       & $P_n$ & $\left<N_{\rm W}\right>_n$ 
       & $P^{7(\rm id)}_n(0.25)$ &    $P^{7(\rm id)}_n(1)$ &    $P^{7(\rm id)}_n(20)$ 
\\[6pt] \hline \\[-6pt]
0  &  $0.9677$ & $6.94$  & $0.9660$ & $6.77$ & $0.9657$       & $0.8706$       & $0.0948$    \\
1  &  $0.0319$ & $6.63$  & $0.0335$ & $6.48$ & $0.0338$       & $0.1219$       & $0.2656$   \\
2  &  $0.0004$ & $6.37$  & $0.0005$ & $6.11$ & $5.1\times 10^{-4}$ & $7.3\times 10^{-3}$ & $0.3187$   \\
3  &  $0$      & ---     & $0$      & ---    & $4.2\times 10^{-6}$ & $2.4\times 10^{-4}$ & $0.2124$   \\
4  &  $0$      & ---     & $0$      & ---    & $2.1\times 10^{-8}$ & $4.9\times 10^{-6}$ & $0.0850$
  \\[6pt] \hline\hline
  \end{tabular}
\end{table*}

Very recently Lee and van der Vegt have published a series of molecular 
dynamics simulations of TBA/water mixtures with varying composition 
\cite{Lee:2005}. In their study, using the SPC model \cite{Berendsen:81} 
for water, they found the original
united-atom OPLS potential \cite{Jorgensen:86} for TBA less
satisfactory and hence derived an improved set of model 
parameters for tertiary butanol, basically by adjusting the
charges on the OH-group. Lee and van der Vegt report a 
too strong association of the TBA molecules for the original 
united-atom OPLS model. 
We would like to point out that reported effect might be also
partly attributed to the use of the SPC water model,
since an analogous solvent-model dependence 
has been recently observed for
simple solutes (noble gases). 
Paschek \cite{Paschek:2004:1} found for the SPC model a substantially
stronger hydrophobic association, which experiences
also a significantly smaller
temperature dependence. For our present study we choose 
the SPCE model on purpose, since it reproduces
several water properties somewhat better than SPC, 
such as OO-radial distribution functions \cite{Hura:2000,Sorenson:2000}, 
the self-diffusion coefficients \cite{vanderSpoel:1998}, as well as the solvation entropy of 
small hydrophobic particles \cite{Paschek:2004:1}.
However, using their models for an aqueous salt solution, preliminary data of
Lee and van der Vegt indicate an association behavior
of TBA very much similar to our MD simulations: An enhanced hydrophobic TBA-TBA
association upon addition of sodium chloride \cite{vegt_lmc06}.

\subsection{A combinatorial picture of TBA/water-content equilibrium in the chloride solvation shell}

In this section we would like to elucidate the mechanism that
determines the composition of the chloride
solvation shell by putting forward a simplified 
picture based on purely combinatorial arguments.
This analysis will provide us 
also an estimate of the
relative difference in binding free energy between 
TBA and water found in our MD simulation, as well as the one 
that would be necessary to create 
the Bowron-Finnney scenario of Chloride-bridged TBA-pairs.

First of all we have statistically analyzed the structure of the first solvation shell of a
chloride ion as it is provided by our MD simulations. 
We have calculated the probability of
finding a solvation shell with no TBA, one TBA molecule, two TBA molecules and
so forth. As members of the first hydration shell are considered all TBA
molecules with $r_{\rm OH-Cl}\leq 0.39\,\mbox{nm}$ and all 
water molecules with
 $r_{\rm OW-Cl}\leq 0.39\,\mbox{nm}$ (see Table \ref{tab:COOR} and Figure \ref{fig:02}).
These data are shown in the left 
part of Table \ref{tab:combinatoric}. Given are results for MD simulations 
at two different salt concentrations. 
Moreover, we also provide the average number of water
molecules $\left<N_{\rm W}\right>_n$ in a solvation shell 
accommodating $n$ TBA molecules. 

The Cl-OW coordination number shown in Table \ref{tab:COOR}, as well as
the values for $\left<N_{\rm W}\right>_n$ given in Table \ref{tab:combinatoric}
indicate that on average about seven to eight molecules 
(water and TBA) form the 
solvation shell of a chloride ion in the simulated solution. Now we would like
to apply
a simplified model of the 
solvation shell. Let us assume we have $m$ sites
(in our case: say seven) in the shell that can be either occupied by a 
water, or a TBA molecule. Then the probability that a site is
occupied by a TBA molecule is according to
\begin{equation}
p_{\rm T} = \frac{K_{\rm eq}\,N_{\rm T}}{K_{\rm eq}\,N_{\rm T}+N_{\rm W}}\;,
\label{eq:Langmuir}
\end{equation}
where $N_{\rm T}$ and $N_{\rm W}$ specify the number of TBA  and water 
molecules in the simulation box, respectively. $K_{\rm eq}$ represents
the constant describing the TBA/water equilibrium per solvation shell site.
In fact, Equation \ref{eq:Langmuir} represents the saturation limit for a
Langmuir-type adsorption 
of a binary mixture (see Ref. \cite{Ben-Naim:StatMech} p. 80).
Correspondingly,
$\Delta G\!=\!-RT \,\ln K_{\rm eq}$ represents the relative free energy change when
replacing one water molecule by a TBA, and might be considered as the
difference in ``binding strength''.
Given there are no further inter-site correlations within the solvation shell, 
the probability to find a chloride ion surrounded by exactly $n$ TBA molecules
follows a binomial distribution, and is given by
\begin{equation}
P^{m(\rm id)}_n\left(K_{\rm eq}\right) = 
  {m \choose n}\;p_{\rm T}^{n}\,(1-p_{\rm T})^{(m-n)} \,.
\end{equation}
On the right hand side of Table \ref{tab:combinatoric} we illustrate
three different cases: For the case of equal binding strength for water and TBA 
($K_{\rm eq}\!=\!1$), the probabilities $P_n$ are just determined by the
composition of the solution (the $N_{\rm T}/N_{\rm W}$ ratio). We would
like to emphasize that even for this case the probability to find 
Cl-bridged dimers is below one percent. When comparing with data
obtained directly from the MD simulation, we find them 
well reproduced assuming $K_{\rm eq}\!\approx\!0.25$. 
This indicates that a TBA-molecule is on average 
more weakly bound to a chloride-ion than a water molecule
by about  $\Delta G\!\approx\!3.4\,\mbox{kJ}\,\mbox{mol}^{-1}$. 
We would like to 
point out that due to the small value of $p_{\rm T}\!\approx\!5\times 10^{-3}$,
the obtained probabilities $P_n$ depend only very weakly
on the variation of the number of sites $m$ (say between $6$ and $9$)
and are well approximated by a Poisson distribution.
At last we would like to discuss a case, where the maximum in $P_n$ is found
for the case of Cl-bridged TBA-dimers ($n\!=\!2$). To generate such a scenario,
the equilibrium has to be shifted largely to the TBA-side with
$K_{\rm eq}\!\approx\!20$, leading to a roughly $80$ times stronger binding
of TBA compared to our MD-simulations, or an increase of about
$11\,\mbox{kJ}\,\mbox{mol}^{-1}$ on a free energy scale.
Concluding, we would like to emphasize that unless
the ion-TBA binding is not substantially stronger than the ion-water
interaction, simply the vast majority of water molecules in
the solution will 
prevent the formation of Cl-bridged TBA-dimers.
Finally, although our employed  model potential is certainly far from 
being perfect, there is actually no indication on what could 
cause such a strong specific chloride-HO interaction. 
Moreover, the variation of the partial
charges on the hydroxyl hydrogen (HO) varies only slightly
among different forcefields:
$+0.418 |\mbox{e}|$ for the all-atom OPLS model \cite{Jorgensen:96},
$+0.423 |\mbox{e}|$ for van der Vegt's optimized united-atom OPLS model
for TBA \cite{Lee:2005}, and $+0.435 |\mbox{e}|$ for
the united-atom TraPPE-forcefield by Chen et al. \cite{Chen:2001}.
A specifically {\em linearly} chloride-bridged TBA pair
would represent a state of additional configurational
order and hence even lower entropy.
Favoring this particular configuration among all others
would therefore require an increasingly lower energy for 
this type of complex.

\subsection{TBA-clustering and ``salting out''}

\begin{figure}[!t]
  \centering
  \includegraphics[angle=0,width=5.0cm]{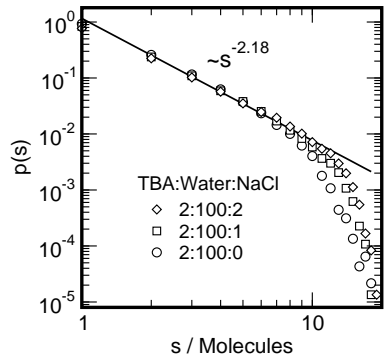}
  \caption{Log-log plot of the probability $p(s)$
    to find a TBA-cluster of size
    $s$, where $s$ indicates the number of TBA-molecules in the cluster.
    The solid line indicates the behavior expected for random bond percolation
    on a 3D-lattice close to the percolation transition \cite{Stauffer}.
  }
  \label{fig:cluster1}
\end{figure}

To determine whether the observed enhanced 
hydrophobic aggregation upon
addition of salt leads to
a ``salting out'' of the TBA-molecules, we  present a cluster size
analysis of the TBA-aggregates. Such an approach has been
recently put forward to study 
the onset of a demixing transition for the case of
binary water-tetrahydrofuran mixtures \cite{Oleinikova:2002}
and to structurally characterize 
aqueous methanol solutions \cite{Dougan:2004}.
Based on the TBA-TBA central carbon pair correlation
function (shown in Figure \ref{fig:01}) we consider two TBA
molecules as ``bound'' when their central carbon C-C distance
is smaller than $0.72\,\mbox{nm}$.
We would like to point out that also hydrogen bonded TBA-TBA
pairs are included by this definition. Based on that definition of
intermolecular connectivity, we identify clusters
of associated molecules. Figure \ref{fig:cluster1}
shows the log-log plot of the  probability to find a cluster of
a certain size $s$ (number of TBA molecules) 
for the aqueous TBA solutions
with increasing
salt concentration. First of all we would like to point out
that for all shown distributions
and for small cluster-sizes, the distribution approaches
$p(s)\sim s^{-\tau}$ with an exponent
of $\tau\approx2.18$, as observed for the case of random bond percolation
on a 3D-lattice close to the percolation transition \cite{Stauffer}. 
This seems to be well in
accordance with the picture
of an equilibrium of randomly associating 
and dissociating clusters of molecules.
For larger cluster sizes, however, we observe a systematic
deviation from the power-law behavior, 
which is due to the constraint of a finite number of 
TBA-molecules in our simulation.
Please note that with increasing salt concentration an increased
number of large clusters appears. This is a direct evidence for
a ``salting out'' effect upon increased salt concentration.
We would like to stress the observation,
that also by visual inspection
the solutions still appear to be 
homogeneous and are apparently situated below a
possible phase separation.
\begin{figure}[!t]
  \centering
  \includegraphics[angle=0,width=5.0cm]{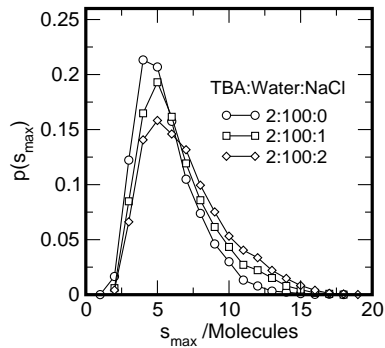}
  \caption{
    Distribution of the size of the largest clusters $s_{\rm max}$ (number of
    TBA molecules) per configuration
    as a function of salt concentration.
    }
  \label{fig:cluster2}
\end{figure}
Moreover, for a system close to a phase separation, we would expect 
the appearance of a ``hump'' at large cluster-sizes.
The relation of the build-up of such a narrow
distribution of cluster-sizes showing a maximum close to the largest
possible cluster size (in our simulation: 20 TBA molecules)
with the onset of phase separation has been
recently demonstrated for the case of tetrahydrofuran/water
mixtures by Oleinikova et al.\cite{Oleinikova:2002}.
However, we cannot fully rule out that the tendency of the system to
phase separate might be also suppressed by finite size effects.
To further substantiate the apparent ``salting out'' tendency,
we show in Figure \ref{fig:cluster2} the 
distribution  of cluster-sizes of the largest
cluster detected in each configuration. The
maximum of each of the distribution functions is located in the vicinity
of 5 molecules. However, there is a clear shift towards larger cluster-sizes
with increasing salt concentration. By promoting larger cluster-sizes
with increasing NaCl concentration,
the TBA-clustering analysis clearly
indicates a ``salting out'' tendency of the
aqueous TBA-solutions.

\subsection{Self-diffusion coefficients}

\begin{table*}[!t]
\caption{
  Self diffusion coefficients $D_{\rm self}$ as obtained from the MD simulations
  for the TBA/Water- and TBA/Water/Salt-
  mixtures and for pure water.
  The star$^*$
  indicates the parameter set for NaCl according to Koneshan et
  al. \cite{Koneshan:1998}.
}
\label{tab:Dself}
\centering
  \ifpretty
  \renewcommand{\arraystretch}{1.1}
  \renewcommand{\tabcolsep}{0.62cm}
  \else
  \renewcommand{\arraystretch}{1.1}
  \renewcommand{\tabcolsep}{0.2cm}
  \fi
  \small
  \begin{tabular}{lccccc} \hline\hline \\[-6pt]
 ~ & \multicolumn{5}{c}{$D/10^{-9}\;\mbox{m}^2\mbox{s}^{-1}$} \\[6pt]
 ~& \multicolumn{5}{c}{TBA:Water:NaCl} \\[6pt]
  Particle & 2:100:1$^*$ & 2:100:2 & 2:100:1 & 2:100:0 &  0:100:0
\\[6pt] \hline \\[-6pt]
$\mbox{H}_2\mbox{O}$ & $2.035\pm0.003$  & $1.975\pm0.005$ & $2.075\pm0.005$  & $2.197\pm0.006$  &  $2.642\pm0.012$ \\
TBA                  & $0.751\pm0.009$  & $0.707\pm0.009$ & $0.749\pm0.011$  & $0.840\pm0.015$  &  ~ \\
$\mbox{Na}^+$        & $0.987\pm0.018$  & $1.041\pm0.013$ & $1.091\pm0.017$  &       ~          &  ~ \\
$\mbox{Cl}^-$        & $1.200\pm0.021$  & $1.162\pm0.014$ & $1.290\pm0.020$  &       ~          &  ~
\\[6pt] \hline\hline
  \end{tabular}
\end{table*}
\begin{table*}[!t]
\caption{
  Parameters characterizing the full
  H-H-dipolar correlation function (according to Eq.\ref{eq:dipolcor}) 
  for the TBA aliphatic hydrogen nuclei
  as obtained from the MD-simulations. Given are both intra- and intermolecular
  contributions.
  The values indicate a by 20 \% increased intermolecular relaxation
  rate $(T_1^{-1})_{\rm inter}$ in the presence of NaCl, whereas the
  intramolecular rate changes only by about 2 \%.
  To obtain the correlation times $\tau_2$ and
  prefactors $\left<r_{\rm HH}^{-6}\right>\!=\!G_2(0)$ 
  (see Eq.\ref{eq:ddcorel}), 
  the correlation function
  was integrated numerically while the tail of the correlation
  function was fitted to a single exponential function (inter: between 120 and 200
  ps; intra: between 30 and 50 ps).
  The corresponding intra- and intermolecular relaxation times were finally
  obtained from Eq.\ref{eq:Westlund:1987}.
}
\label{tab:corel}
  \centering
  \ifpretty
  \renewcommand{\tabcolsep}{0.63cm}
  \renewcommand{\arraystretch}{1.0}
  \else
  \renewcommand{\tabcolsep}{0.2cm}
  \renewcommand{\arraystretch}{1.0}
  \fi
 \begin{tabular}{lcccccc} \hline\hline \\[-4pt]
    TBA:Water:NaCl &
   \multicolumn{3}{c}{inter} &
   \multicolumn{3}{c}{intra} \\[0.5em]
   ~ &
   $\left<r_{\rm HH}^{-6}\right> / \mbox{nm}^{-6}$ &
   $\tau_2 / \mbox{ps}$  &
   $T_1 / \mbox{s}$ &
   $\left<r_{\rm HH}^{-6}\right> / \mbox{nm}^{-6}$ &
   $\tau_2 / \mbox{ps}$ &
   $T_1 / \mbox{s}$
\\[6pt] \hline \\[-6pt]
   2 : 100         & 1667 &  8.89  &  79.0 & 76920 & 4.41 & 3.45 \\
   2 : 100 : 1     & 1739 & 10.25  &  65.7 & 74242 & 4.67 & 3.38 \\
   2 : 100 : 2     & 1958 & 10.61  &  56.3 & 74239 & 4.85 & 3.25 \\
   2 : 100 : 1$^*$ & 1681 & 10.27  &  67.8 & 74249 & 4.68 & 3.37 
  \\[6pt] \hline\hline
  \end{tabular}
\end{table*}
We have determined the self diffusion coefficients for all
particle types in the MD simulation 
from the mean square displacement 
according to
\begin{equation}
D=\frac{1}{6} \lim_{t\rightarrow\infty}
\frac{\partial}{\partial t}
\left<
|\vec{c}(t) - \vec{c}(0)|^2
\right>\;,
\end{equation}
where $\vec{c}$ represents the position of the molecules center of mass.
In fact, the diffusion coefficients are obtained from the
slope of the mean square displacement over the 
time-interval between $100\,\mbox{ps}$ and $500\,\mbox{ps}$.
The lower boundary has been chosen to be large compared to the
the average intermolecular association times. 
The diffusion coefficients obtained from the simulations
are given in Table \ref{tab:Dself}. In addition, 
the experimental self-diffusion coefficients for TBA-d1 in
in heavy water/salt solutions are shown in Table \ref{tab:EXPT}.

It is evident that the experimental diffusion coefficients
of TBA are substantially smaller than the values obtained from MD simulations.
This has to be largely 
attributed to the fact that
heavy water is used as a solvent in the experiment, 
which has a significantly smaller diffusion
coefficient ($D\!=\!1.768\times 10^{-9}\;\mbox{m}^2\,\mbox{s}^{-1}$ at 
$298.25\,\mbox{K}$ \cite{Price:2000}) compared to $\mbox{H}_2\mbox{O}$
($D\!=\!2.299\times 10^{-9}\;\mbox{m}^2\,\mbox{s}^{-1}$ at 
$298.2\,\mbox{K}$ \cite{Mills1973,Holz:2000}). However, even
when taking this effect into account, the diffusion coefficient 
of TBA according to the MD simulations seems to be 
overestimated by about $40\%$, which
could indicate an increased number of
associated TBA-TBA 
pairs in the experiment. This discrepancy may at least partly
also be attributed
to a possible enhanced ``retardation effect'' \cite{Haselmeier:95}
of heavy water in the hydrophobic hydration shell of TBA. The
``retardation effect'' is clearly seen in the simulations as 
the slowing down of the water molecules in the TBA/water solution.
However, given that the water retardation effect is related
to the structuring of water in the hydrophobic hydration shell, it
should rather be weaker in SPCE than in real water, 
since SPCE has been shown to underestimate the solvation
entropy of hydrophobic particles \cite{Paschek:2004:1}.

In addition, we find that the decrease of the TBA diffusion coefficient 
upon addition of salt
by $6.4\%$ per $\mbox{mol}\,\mbox{l}^{-1}$ NaCl is significantly
smaller than by the $15.8\%$ per $\mbox{mol}\,\mbox{l}^{-1}$ NaCl as obtained
in the MD simulation. The overestimated salt-effect
might be attributed to the potentially too small 
diffusion coefficients observed for $\mbox{Na}^+$ and particularly
$\mbox{Cl}^-$. Scaling the ion self-diffusion coefficients
with the diffusion coefficient of water in
the TBA/water system (as approximation), one would expect
diffusion coefficients of 
$D(\mbox{Na}^+)\!\approx\!1.18\times 10^{-9}\;\mbox{m}\,\mbox{s}^{-2}$ and
$D(\mbox{Cl}^-)\!\approx\!1.70\times 10^{-9}\;\mbox{m}\,\mbox{s}^{-2}$
for the $c(\mbox{NaCl})\!=\!1.0\;\mbox{mol}\,\mbox{l}^{-1}$ concentration
(using diffusion coefficients 
from Ref. \cite{MillsLobo} 
for the ions in aqueous solutions at the 
given concentration for $25^\circ\mbox{C}$).
Although the interaction with the TBA molecules might also
have a non-negligible effect, the MD simulations seem to 
underestimate the self-diffusion 
coefficients of the ions by about $13\%$ and $46\%$, respectively. 
To confirm this observation we have additionally 
calculated self-diffusion coefficients for a purely aqueous 
salt solution
containing 500 SPCE molecules and 16 ion pairs at $298\,\mbox{K}$
and $1\,\mbox{bar}$,
with a concentration of $1.67\;\mbox{mol}$. We obtain diffusion coefficients
of 
$D(\mbox{Na}^+)\!=\!1.07\times 10^{-9}\;\mbox{m}\,\mbox{s}^{-2}$,
$D(\mbox{Cl}^-)\!=\!1.21\times 10^{-9}\;\mbox{m}\,\mbox{s}^{-2}$,  and
$D(\mbox{H}_2\mbox{O})\!=\!2.18\times 10^{-9}\;\mbox{m}\,\mbox{s}^{-2}$.

The diffusion coefficients of the ions clearly indicate that 
there is a need for improvement of the employed ion parameter sets.
Particularly the hydration-strength of the chloride ion seems to be
overestimated by the present models. The overall agreement with the
n-scattering data and the less satisfactory diffusion data, however,
might just reflect the more pronounced sensitivity of kinetic
quantities on details of the pair correlation function, such as
the depth of the first minimum in the ion-water pair
correlation function. This property is critically related to
the water-exchange rate.

\subsection{TBA-TBA association and the A-parameter }

\begin{figure}[!t]
  \centering
  \includegraphics[angle=0,width=5.0cm]{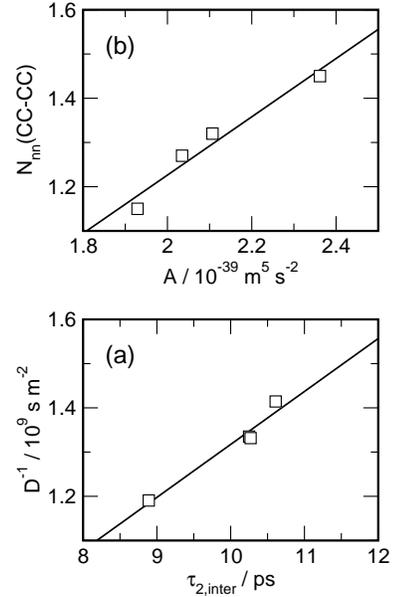}
  \caption{
    $A$-parameter related-quantities as obtained from
    the MD-simulation:
    a) Inverse self-diffusion coefficient of TBA $D^{-1}$
    versus the intermolecular dipolar correlation time $\tau_{2,\rm inter}$ 
    of the aliphatic protons in TBA 
    for the different MD simulations.
    b) TBA-TBA aliphatic carbon coordination number versus
    the $A$-parameter.
    }
  \label{fig:DtauT1N}
\end{figure}
\begin{figure}[!t]
  \centering
  \includegraphics[angle=0,width=5.0cm]{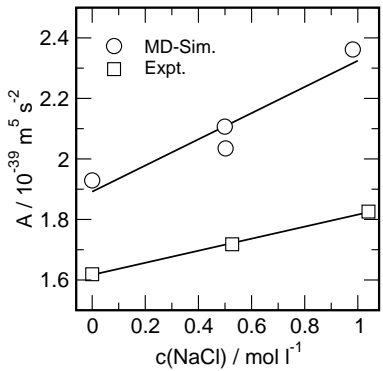}
  \caption{
    $A$-parameter obtained from MD simulation and experiment
    as a function of salt concentration.
    }
  \label{fig:Aparam}
\end{figure}
Given the uncertainties associated with the use of
empirical forcefield-models,
we have tried to find out, whether there is also
experimental evidence 
for an enhanced  hydrophobic association scenario, as it is
suggested by our MD simulations.
Therefore we perform nuclear magnetic
relaxation experiments on aqueous solutions of tertiary butanol
with varying salt concentration.
Given a non-changing
intermolecular dipolar correlation time $\tau_{2,\rm inter}$, an increasing
relaxation rate of the methyl hydrogen nuclei
would {\em directly} indicate an
association behavior as observed
in our MD-simulations. However,
a decreasing relaxation rate would support the scenario
obtained by Bowron and Finney. Since the correlation
times are likely to be changing, we 
follow the $A$-parameter approach proposed
by Hertz and co-workers, and discussed extensively in a previous section.
The $A$-parameter approach is based on the assumption of a linear relationship
between the intermolecular dipolar correlation time $\tau_{2,\rm inter}$
and the inverse self-diffusion coefficient of the solute molecules 
(Equation \ref{eq:A_approx}),
which can be both obtained independently. To prove this
we have calculated the full inter- and intramolecular dipole-dipole correlation
functions according to Eq. \ref{eq:dipolcor}
for the aliphatic hydrogens for all MD-simulations,
assuming all other protons to be exchanged by deuterons. Here
only like-spin $^1\mbox{H}-^1\mbox{H}$ dipolar interactions are considered.
A quantitative description of the intra- and intermolecular contributions
as well as the calculated relaxation times $T_1$ 
are given in Table \ref{tab:corel}. The TBA self-diffusion coefficients
for each of the simulated systems is given in Table \ref{tab:Dself}.
Figure \ref{fig:DtauT1N}a shows that the 
intermolecular correlation times
$\tau_{2,\rm inter}$ and the inverse self diffusion coefficient
of TBA are indeed almost linearly related, establishing
the $A$-parameter approach as a valid approximation
in the present case.
Moreover, as Figure \ref{fig:DtauT1N}b demonstrates, 
the $A$-parameter from the MD-data is also almost linearly related
to the TBA-TBA coordination number, provoking an
interpretation of the $A$-parameter as an approximate direct 
measure of the TBA-TBA coordination number. 

The $A$-parameters obtained experimentally 
from the measured diffusion coefficients and relaxation rates
are summarized in Table \ref{tab:EXPT}. Both experimental and simulated
$A$-Parameters are shown as well in Figure \ref{fig:Aparam}.
MD-simulation and experiment agree at least in a qualitative sense:
with increasing salt concentration the rising $A$-parameter indicates enhanced
hydrophobic TBA-TBA aggregation. The quantitative difference 
of about $25\%$ between the experimental and simulated $A$-parameters
might be partially attributed to imprecisions of the 
employed potential model, but might be based as well to a certain degree on
the necessity for using $\mbox{D}_2\mbox{O}$ as a solvent in
the NMR-measurements.
Making use of the almost linear relationship between the $A$-parameter
and the TBA-TBA coordination number, we would finally  like to
determine approximate TBA-TBA
coordination numbers for the NMR experimental data.
Doing so, we get coordination numbers of 
$N_{nn}(TBA)\!=0.975$, $1.040$, and $1.111$ for the solutions
with salt concentrations $0.0\,\mbox{mol}\,\mbox{l}^{-1}$,
$0.5270\,\mbox{mol}\,\mbox{l}^{-1}$, and
$ 1.0403 \,\mbox{mol}\,\mbox{l}^{-1}$, respectively.
Although by about $25\%$ smaller than the coordination numbers
obtained from MD-simulation, and suggesting a smaller concentration
variation, the experimentally obtained $A$-parameters and 
coordination numbers still tend to confirm an enhanced hydrophobic 
aggregation of the TBA molecules with increasing salt concentration.
Finally, we would like to point out that these ``experimental'' 
coordination numbers are also in agreement with the errorbars
of the n-scattering data shown in Table \ref{tab:COOR}.

\section{CONCLUSIONS}

We have used a combination of molecular dynamics simulations and 
nuclear magnetic relaxation measurements to investigate
the effect of salt (sodium chloride) on the association-behavior
of tertiary butanol molecules in an aqueous solution.
We have shown that the application of
the so-called ``$A$-parameter''-approach,
proposed by Hertz and co-workers, employing 
solute-solute intermolecular 
$^1H$-relaxation times and solute diffusion coefficients
to determine the (relative) degree of association of solute
molecules in aqueous solution, is
well justified in the present case.
Moreover, our MD-simulations establish
an almost linear relationship between the 
$A$-parameter
and  the TBA-TBA coordination number. 

Both MD-simulations and NMR-experiment tend to
support a classical hydration and ``salting out''
picture of an enhanced tendency of forming hydrophobic contacts 
between the TBA-molecules in the presence
of salt. An increasing salt concentration is hence
found to strengthen the solute-solute hydrophobic interaction.
Consequently, also the TBA-cluster-size distributions reveal a growing size 
of the TBA-aggregates as the salt concentration increases and
therefore show directly 
the ``salting out'' tendency.
Apparently, the TBA molecules behave closely similar to
purely hydrophobic solutes, as recently shown
by Ghosh et al.\cite{Ghosh:2003} for the case of
hydrophobic methane particles dissolved in aqueous salt solutions.
In the light of the scenario recently suggested 
by Koga and Widom \cite{Widom:2003,Koga:2004}, the salt 
thus would lead to an increase of
the excess chemical potential 
of the hydrophobic groups (reduce their solubility) and would
therefore provoke an enhanced aggregation.
Based on lattice model calculations Koga and Widom have proposed
an almost inverse
linear relationship between the excess chemical potential of
hydrophobic solutes and their hydrophobic interaction strength.

Our detailed structural analysis of the simulation data does 
not provide any evidence for the presence of
chloride-bridged butanol-pairs, as proposed 
by Bowron and Finney \cite{Bowron:2002:1,Bowron:2003}.
Moreover, a combinatorial analysis of the composition
of the chloride solvation shell reveals that the
formation of a significant amount
of dimers is probably just 
prevented by the vast majority of water molecules
in the solution.
Finally, we would like to emphasize that although our results suggest a structurally 
completely distinct scenario,
the molecular dynamics simulations, as well as the coordination numbers
obtained indirectly from our NMR-experimental data,
appear to be largely within the 
experimental errorbars of the n-scattering data 
and therefore seem to be compatible with
Bowron and Finney work \cite{Bowron:2003}.

\section*{Acknowledgments}

We acknowledge support from the Deutsche Forschungsgemeinschaft
(FOR-436 and GK-298) and the University of Dortmund 
(``Forschungsband Molekulare Aspekte der Biowissenschaften'').
The authors are grateful to M. Holz for critically reading
the manuscript.

\end{document}